 \documentclass[final,5p,times,twocolumn,authoryear]{elsarticle}

\usepackage{amssymb}
\usepackage{lipsum}
\usepackage{url}
\usepackage{amsmath}

\usepackage{xcolor}

\usepackage{lineno}

\journal{Journal of High Energy Astrophysics}

\begin{document}

\begin{frontmatter}



\title{Data-Driven Constraints on Magnetar Population: No Evidence for a Distinct White Dwarf Channel}

\author[1]{R. V. Lobato}
\ead{lobato@cbpf.br}

\affiliation[1]{organization={Centro Brasileiro de Pesquisas F\'isicas},
addressline={Rua Dr. Xavier Sigaud 150},
city={Rio de Janeiro},
postcode={22290-180},
state={RJ},
country={Brazil}}

\begin{abstract}
Magnetars are usually interpreted as highly magnetized neutron stars, yet a small subset of low spin-down sources has motivated alternative scenarios involving highly magnetized white dwarfs.
We test whether the observed magnetar sample is consistent with a single neutron-star population or whether the data favor an additional compact-object channel.
We combine exploratory machine-learning diagnostics with hierarchical Bayesian population modeling. First, we apply K-means clustering and principal component analysis for visualization in a five-dimensional feature space $(P,\dot{P},L_X,kT,|Z|)$, where $P$ is the spin period, $\dot{P}$ its time derivative, $L_X$ the X-ray luminosity, $kT$ the thermal spectral temperature, and $|Z|$ the absolute Galactic scale height, and then train a Random Forest classifier with leave-one-out cross-validation to identify the observables driving the empirical split. We subsequently construct a hierarchical Bayesian mixture model that links spin parameters to magnetic-field distributions through covariate-dependent mixing fractions. Posterior inference is performed with Hamiltonian Monte Carlo, and predictive performance is assessed with Pareto-smoothed importance sampling leave-one-out cross-validation.
The exploratory analysis reveals a reproducible sub-structure: the Random Forest reaches $>95\%$ LOOCV accuracy, with $L_X$, $\dot{P}$, and $kT$ emerging as the dominant predictors. However, the Bayesian comparison shows no statistically significant preference for a two-population model. Instead, a few low spin-down sources receive intermediate posterior membership probabilities, indicating that they are better interpreted as transitional or outlying objects than as members of a clearly distinct class.
Overall, current data do not require a separate white-dwarf magnetar population. The main result is therefore conservative but strong: the observed sample is adequately described by a predominantly neutron-star population, while still allowing physically interesting deviations in specific sources.
\end{abstract}




\end{frontmatter}



\section{Introduction}
\label{sec:1}

Magnetars are generally understood as neutron stars endowed with ultra-strong magnetic fields of order $10^{14}$-$10^{15}$ G, powering their high-energy emission through magnetic energy dissipation rather than rotational energy loss \citep{mereghetti/2008, turolla/2015, mereghetti/2015, kaspi/2017, esposito/2020}. Observationally, they are identified through their long spin periods, large spin-down rates, and characteristic X-ray emission properties. The currently known population is compiled in the McGill Magnetar Catalog \citep{olausen/2014}, which provides a comprehensive dataset of timing, spectral, and environmental properties.

Despite the overall success of the neutron star interpretation, a subset of magnetars exhibits unusually low spin-down rates, leading to inferred dipole magnetic fields significantly below the canonical magnetar range. A notable example is SGR~0418+5729, which has been interpreted as a low-magnetic-field magnetar \citep{rea/2010, rea/2013}. SGR~0418+5729 is a well-known outlier because its inferred dipole field is only $B \approx 6 \times 10^{12}$~G, well below the canonical magnetar range. Similarly, Swift~J1822.3$-$1606 is often discussed as a low-field magnetar candidate \citep{rea/2012}, with an inferred dipole field of order $\sim 10^{13}$~G, still below typical magnetar values. These objects challenge the standard picture and raise the possibility that additional physical mechanisms, such as magnetic field burial, decay, or alternative compact object scenarios, may play a role in shaping the observed phenomenology.

One such alternative proposes that some magnetar-like sources could instead be highly magnetized, rapidly rotating white dwarfs \citep{malheiro/2012, coelho/2014, lobato/2016a, mukhopadhyay/2016}. In this scenario, the scaling between spin parameters and magnetic field strength differs from that of neutron stars due to the distinct stellar structure, particularly the larger radius and moment of inertia. Because the viability of this interpretation remains open, we treat it as a testable hypothesis and perform a systematic statistical comparison between competing population models.

Motivated by this unresolved interpretation, we use a hierarchical Bayesian framework to test, rather than assume, whether the observed magnetar population is adequately described by a single neutron star population or is better represented by a mixture that allows an alternative compact-object channel. By modeling the relationship between spin parameters and magnetic field distributions, and incorporating observational covariates, we treat population heterogeneity as a hypothesis to be evaluated with the data. This strategy is aligned with recent machine-learning and Bayesian inference developments in astronomy, gravitational waves, neutron-star and dense-matter studies \citep{lobato/2022, lobato/2022b, chimanski/2023, fujimoto/2024, zhou/2024, stergioulas/2024-mlgw, papigkiotis/2025, li/2025-prc, li/2025-likelihood, dong/2025, ng/2025}. Model comparison is performed using modern Bayesian techniques, including Pareto-smoothed importance sampling leave-one-out cross-validation \citep{vehtari/2017}, allowing for a robust evaluation of predictive performance.

Our approach provides a unified statistical approach to quantify deviations from the canonical magnetar model and to identify objects that may occupy transitional regimes within the population.

\section{Exploratory Data Analysis \& Covariate Selection}
\label{exploration}

Before introducing the Bayesian hierarchy model \citep{gelman/1995} it was performed an data-driven exploration, we tested whether the observed sample exhibits natural structure without imposing any physical labels. We applied an unsupervised pipeline using the period of the source $P$, the rate that this period change, i.e., the spin-down rate $\dot{P}$, the X-ray luminosity of the source $L_X$, the surface temperature $kT$, and the absolute Galactic scale height $|Z|$: first standardizing the variables in log-space. We applied the $K$-means algorithm ($K=2$) directly to this 5-dimensional standardized feature space to identify population split, then we used principal component analysis (PCA) to project these 5D clusters on the 2D plane capturing the maximum variance\citep{linde/1980, lloyd/1982, halkidi/2001, jolliffe/2013}. In the McGill catalog, $L_X$ typically represents the persistent (quiescent) luminosity, generally integrated over the $2-10$ keV energy band therefore depend on both the assumed spectral model and distance uncertainties. Magnetars distances are derived from highly uncertain proxies such as supernova remmants.

\begin{figure}
   \centering
   \includegraphics[width=\hsize]{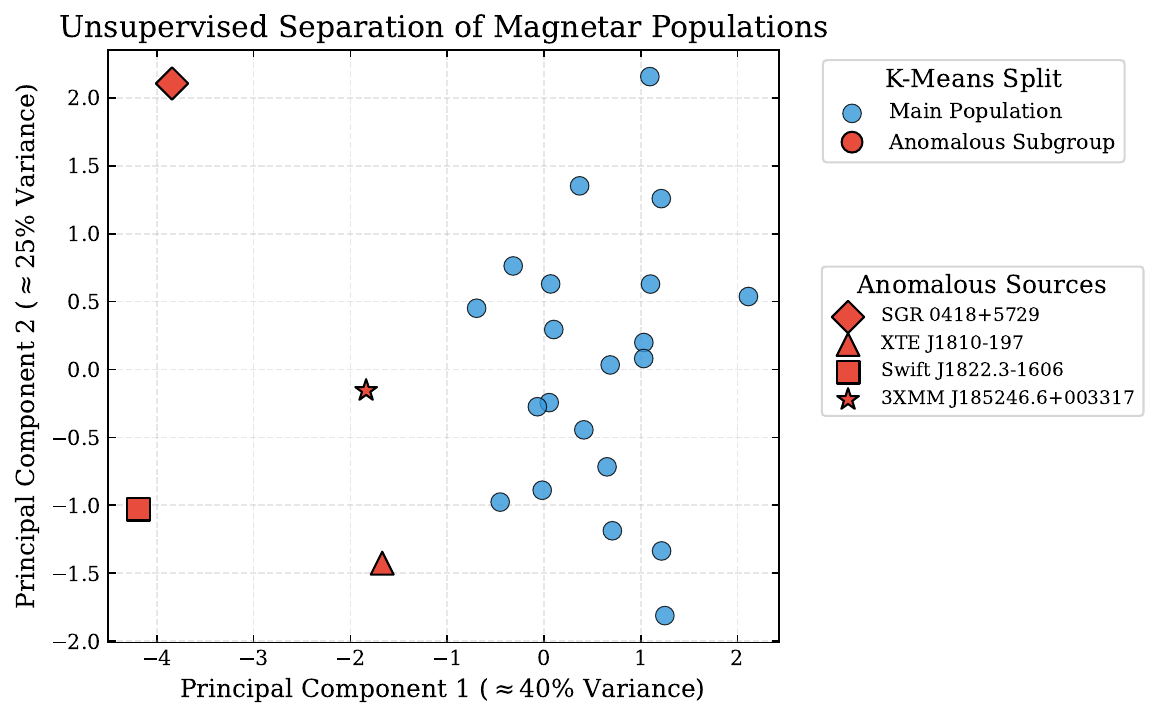}
   \caption{Unsupervised exploratory structure in the magnetar sample using the standardized five-dimensional feature space $(P,\dot{P},L_X,kT,|Z|)$. The sample is partitioned with $K$-means ($K=2$), cluster are then projected with PCA in 2D. The secondary cluster isolates low spin-down, low-$x$ candidates ($x\equiv\log_{10}(P\dot{P})$), motivating a two-component generative model. This indicates an empirical sub-structure that warrants formal mixture testing.}
   \label{fig:eda-pca}
\end{figure}

Figure~\ref{fig:eda-pca} suggests a separation into two groups in the PCA plane, with one compact sub-population containing low spin-down outliers. Importantly, this grouping is recovered independently of any neutron star/white dwarf prior assumptions. Sources such as SGR~0418+5729 and Swift~J1822.3$-$1606 are assigned to, or lie near, this secondary cluster when the algorithm uses only timing and luminosity information.

The empirical separation was validated using the silhoute score metric on the standardized 5D feature space. The maximum score is obtained for $K=2$ (silhouette $\approx 0.39$), while larger values of $K$ lead to significantly lower scores ($\lesssim 0.2$). This indicates a modest preference for a two-group structure.

To identify the physical drivers behind this empirical separation, we trained a Random Forest classifier \citep{ho/1995} on the same 5D standardized feature space to predict the unsupervised $K$-means cluster assignment and evaluated it with Leave-One-Out Cross-Validation (LOOCV) \citep{gelman/2014}. The classifier achieves $>95\%$ accuracy, suggesting that the partition is reproducible from the observed feature space rather than solely an artifact of projection, although this should be interpreted with caution given the sample size. To quantify the relative contribution of each physical parameter to this classification, we compute the Gini importance, also known as Mean Decrease Impurity \citep{breiman/2001, hastie/2009}. In Random Forest, Gini importance measures the total reduction in class impurity (the probability of misclassification) achieved by splitting the data on a specific variable, averaged across all decision trees in the ensemble. Consequently, a higher Gini score indicates that a feature is more fundamentally responsibly for separating the populations.

\begin{figure}
   \centering
   \includegraphics[width=\hsize]{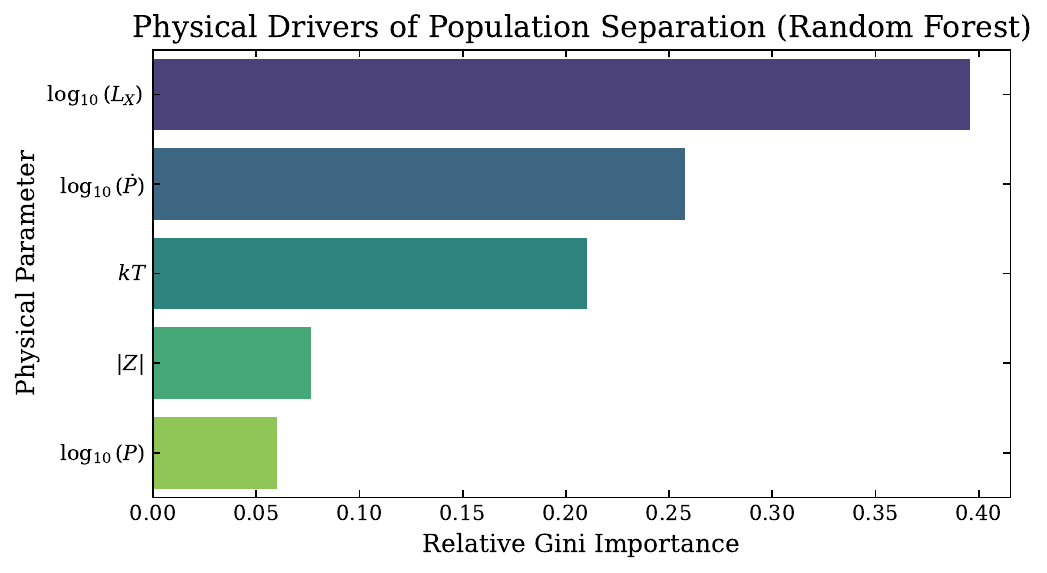}
   \caption{Random Forest feature importance for predicting unsupervised cluster labels under LOOCV. The Gini-importance ranking indicates that $L_X$, $\dot{P}$, and $kT$ are the primary drivers of the observed population split. This justifies using these variables as key covariates in the hierarchical mixture model.}
   \label{fig:eda-rf}
\end{figure}

The Gini-importance summary in Fig.~\ref{fig:eda-rf} shows that $L_X$, $\dot{P}$, and $kT$ dominate the explained variance in cluster membership, with weaker contributions from the remaining covariates. This ranking directly motivates their central role in the subsequent covariate-dependent mixture formulation.

This empirical, label-free partition provides data-driven motivation for adopting a two-component mixture model in the subsequent PyMC analysis \citep{abril-pla/2023}.

\section{Hierarchical Bayesian Mixture Model}

Motivated by the feature-importance ranking derived in Sect.~\ref{exploration}, we construct a hierarchical mixture model in which the probability that source $i$ belongs to the white-dwarf channel is denoted by $\pi_i \equiv p_{\mathrm{WD},i}$. This mixing probability is parameterized as a logistic function of $\log_{10}L_{X,i}$, $kT_i$, and $|Z_i|$.

We do not include $\dot{P}$ as an explicit covariate in the mixing regression, even though it carries strong discriminatory power, because it enters directly through the generative observable
\begin{equation}\label{ppdot}
x_i \equiv \log_{10}(P_i\dot{P}_i).
\end{equation}
In this way, the spin parameters $P_i$ and $\dot{P}_i$ provide the primary physical information in the likelihood, while $L_X$, $kT$, and $|Z|$ act as complementary spectral and environmental covariates. The quantity $x_i$ is directly related to the dipole magnetic field strength under standard spin-down assumptions \citep{mereghetti/2008, turolla/2015, malheiro/2012}, and is used only in the Bayesian model rather than in the exploratory machine-learning stage.

\subsection{Hierarchical Bayesian Model}

We model the population as a two-component mixture of neutron stars (NS) and white dwarfs (WD). For each class $c \in \{\mathrm{NS},\mathrm{WD}\}$, we assume vacuum dipole braking in an effective-field approximation \citep{mereghetti/2008, malheiro/2012, lobato/2016a}, where $B_i$ is the effective dipole magnetic field of source $i$,
\begin{equation}
P_i\dot{P}_i = K_c B_i^2,
\end{equation}
with
\begin{equation}
K_c = \frac{8\pi^2 R_c^6}{3c_{\mathrm{light}}^3 I_c}.
\end{equation}
Here, $R_c$ and $I_c$ are the characteristic radius and moment of inertia of class $c$, respectively, and $c_{\mathrm{light}}$ is the speed of light. In the more general case, an obliquity factor appears, $P_i\dot{P}_i = K_c B_i^2\sin^2\alpha_i$, where $\alpha_i$ is the angle between the spin and magnetic axes; here that dependence is absorbed into the effective dipole field $B_i$.

Using cgs units, we adopt representative fixed structural constants
\begin{align}
I_{\mathrm{NS}} &= 10^{45}\,\mathrm{g\,cm^2}, & R_{\mathrm{NS}} &= 10^6\,\mathrm{cm}\;(10\,\mathrm{km}), \\
I_{\mathrm{WD}} &= 10^{50}\,\mathrm{g\,cm^2}, & R_{\mathrm{WD}} &= 3\times10^8\,\mathrm{cm}\;(3000\,\mathrm{km}),
\end{align}
which imply
\begin{align}
\log_{10} K_{\mathrm{NS}} &\simeq -39.01, \\
\log_{10} K_{\mathrm{WD}} &\simeq -29.15.
\end{align}
These fixed values are simplifying assumptions; realistic equation-of-state and mass-dependent variations in $R$ and $I$ would broaden the inferred field distributions. Numerically, $K_{\mathrm{WD}}/K_{\mathrm{NS}} \simeq 7.3\times10^{9}$. This large offset is primarily a structural effect: since $K\propto R^6/I$, the much larger white-dwarf radius ($R_{\mathrm{WD}}=3000$~km versus $R_{\mathrm{NS}}=10$~km) strongly amplifies $K$ despite the larger $I_{\mathrm{WD}}$. Consequently, for the same observed $x_i$ (equivalently $P_i\dot{P}_i$), the inferred dipole field in the WD channel is smaller than in the NS channel by the factor $\sqrt{K_{\mathrm{NS}}/K_{\mathrm{WD}}}\simeq1.2\times10^{-5}$.

Conditioning on the latent class label $c_i \in \{\mathrm{NS},\mathrm{WD}\}$, Eq.~\eqref{ppdot} becomes

\begin{equation}
x_i = \log_{10} K_{c_i} + 2\log_{10} B_i.
\end{equation}

We model the latent field distribution through class-dependent hyperparameters $\mu_c$ and $\sigma_c$, prior-location and prior-scale parameters $m_c$, $s_c$, and $\tau_c$, and a Student-$t$ degrees-of-freedom parameter $\nu$:

\begin{align}
\Pr(c_i=\mathrm{WD}) &= \pi_i, \qquad \Pr(c_i=\mathrm{NS}) = 1-\pi_i, \\
\log_{10} B_i\,|\,(c_i=c) &\sim \mathcal{N}(\mu_c,\sigma_c^2), \\
\mu_c &\sim \mathcal{N}(m_c,s_c^2), \\
\sigma_c &\sim \mathrm{HalfNormal}(\tau_c), \\
\nu &\sim \mathrm{Exponential}(0.1) + 2.
\end{align}
The hyperparameters $\mu_{\mathrm{NS}}$ and $\mu_{\mathrm{WD}}$ denote the mean logarithmic magnetic-field strengths of the neutron-star and white-dwarf populations, while $\sigma_{\mathrm{NS}}$ and $\sigma_{\mathrm{WD}}$ represent their intrinsic dispersions. The quantities $m_c$ and $s_c$ specify the centers and widths of the Gaussian priors on $\mu_c$, and $\tau_c$ sets the scale of the HalfNormal prior on $\sigma_c$. Weakly informative Normal and HalfNormal priors provide soft regularization: they encode the expectation that the parameters lie in physically plausible neighborhoods, while leaving sufficiently broad tails so that the data can override the priors when needed. This helps the Hamiltonian Monte Carlo (NUTS) sampler avoid unphysical regions of parameter space, especially for positive scale parameters, without preventing the likelihood from dominating the final inference.

The degrees-of-freedom parameter $\nu$ of the Student-$t$ distribution is assigned the prior $\nu \sim \mathrm{Exponential}(0.1) + 2$, which favors moderately heavy-tailed likelihoods while ensuring $\nu>2$ so that the variance remains finite. Weakly informative priors are adopted to reflect current uncertainties while preventing unphysical values and maintaining stable inference. The complete list of priors is summarized in Table~\ref{tab:priors}. In particular, the prior for $\mu_{\mathrm{NS}}$ is centered at $\log_{10} B \sim 14.5$, consistent with typical magnetar field strengths in the range $10^{14}$--$10^{15}$~G reported \citep{turolla/2015, kaspi/2017}. For the white-dwarf hypothesis, the prior for $\mu_{\mathrm{WD}}$ is centered at $\log_{10} B \sim 8$, covering the range of highly magnetized white dwarfs ($10^{6}$--$10^{9}$~G) discussed in the literature \citep{ferrario/2015}, while still allowing more extreme scenarios \citep{malheiro/2012, mukhopadhyay/2016}.

From the linear relation above, it follows that the latent, noise-free predictor $x_i^{\star}$ satisfies

\begin{equation}
x_i^{\star}\,|\,(c_i=c) \sim \mathcal{N}\!\left(\log_{10}K_c + 2\mu_c,\,4\sigma_c^2\right).
\end{equation}

Here, $x_i^{\star}$ denotes the latent (noise-free) value of $x_i$ implied by the class-$c$ field distribution. Equivalently, the latent predictor has location $\log_{10}K_c + 2\mu_c$ and intrinsic standard deviation $2\sigma_c$.

To obtain a likelihood that is robust to outliers, we then replace the Gaussian kernel by a Student-$t$ working likelihood with the same location and with intrinsic and class-specific extra scatter $\sigma_{c,\mathrm{obs}}$ added in quadrature at the scale level:

\begin{equation}
x_i\,|\,(c_i=c) \sim \mathrm{StudentT}\!\left(\nu,\,\log_{10}K_c + 2\mu_c,\,\sqrt{4\sigma_c^2 + \sigma_{c,\mathrm{obs}}^2}\right).
\label{student}
\end{equation}

In the Student-$t$ likelihood above, the third argument is the scale parameter, not the standard deviation; for $\nu>2$, the corresponding variance is $\nu\left(4\sigma_c^2 + \sigma_{c,\mathrm{obs}}^2\right)/(\nu-2)$. The quantities $\sigma_{\mathrm{NS},\mathrm{obs}}$ and $\sigma_{\mathrm{WD},\mathrm{obs}}$ are class-specific extra-scatter parameters (Table~\ref{tab:priors}), not source-by-source propagated measurement uncertainties from $P_i$ and $\dot{P}_i$, and $\pi_i=\Pr(c_i=\mathrm{WD})$. The broader prior adopted for the WD component reflects the larger uncertainty associated with this hypothetical population. In particular, WD candidates correspond to outlying sources with less well-constrained physical interpretations, and may exhibit additional variability or modeling mismatch relative to the neutron-star population. Allowing for increased extra scatter in the WD channel prevents the model from artificially over-constraining these objects.

\begin{table*}
    \begin{center}
    \caption{Summary of Prior Distributions for the Hierarchical Mixture Model.}
    \vspace{3mm}
    \begin{tabular}{ccl}
    \hline
    \multicolumn{1}{c|}{{Parameter}} & \multicolumn{1}{c|}{{Prior Distribution}} & \multicolumn{1}{c}{{Physical Interpretation}} \\ \hline
    \multicolumn{1}{c|}{$\mu_{\rm NS}$} & \multicolumn{1}{c|}{$\mathcal{N}(14.5, 1.0)$} & \multicolumn{1}{l}{Mean $\log_{10}(B)$ for the Neutron Star channel} \\
    \multicolumn{1}{c|}{$\mu_{\rm WD}$} & \multicolumn{1}{c|}{$\mathcal{N}(8.0, 2.0)$} & \multicolumn{1}{l}{Mean $\log_{10}(B)$ for the White Dwarf channel} \\
    \multicolumn{1}{c|}{$\sigma_{\rm NS}$} & \multicolumn{1}{c|}{$\text{HalfNormal}(1.0)$} & \multicolumn{1}{l}{Intrinsic $\log_{10}(B)$ scatter for Neutron Stars} \\
    \multicolumn{1}{c|}{$\sigma_{\rm WD}$} & \multicolumn{1}{c|}{$\text{HalfNormal}(1.0)$} & \multicolumn{1}{l}{Intrinsic $\log_{10}(B)$ scatter for White Dwarfs} \\
    \multicolumn{1}{c|}{$\alpha$} & \multicolumn{1}{c|}{$\mathcal{N}(0, 2.0)$} & \multicolumn{1}{l}{Global intercept for WD mixture fraction} \\
    \multicolumn{1}{c|}{$\beta_{L_X}$} & \multicolumn{1}{c|}{$\mathcal{N}(0, 0.5)$} & \multicolumn{1}{l}{Logistic weight for X-ray luminosity} \\
    \multicolumn{1}{c|}{$\beta_{kT}$} & \multicolumn{1}{c|}{$\mathcal{N}(0, 0.5)$} & \multicolumn{1}{l}{Logistic weight for surface temperature} \\
    \multicolumn{1}{c|}{$\beta_{|Z|}$} & \multicolumn{1}{c|}{$\mathcal{N}(0, 0.5)$} & \multicolumn{1}{l}{Logistic weight for Galactic scale height} \\
    \multicolumn{1}{c|}{$\sigma_{\rm NS, obs}$} & \multicolumn{1}{c|}{$\text{HalfNormal}(0.3)$} & \multicolumn{1}{l}{Observation/timing scatter for NS candidates} \\
    \multicolumn{1}{c|}{$\sigma_{\rm WD, obs}$} & \multicolumn{1}{c|}{$\text{HalfNormal}(0.5)$} & \multicolumn{1}{l}{Observation/timing scatter for WD candidates} \\
    \multicolumn{1}{c|}{$\nu$} & \multicolumn{1}{c|}{$\text{Exponential}(0.1) + 2$} & \multicolumn{1}{l}{Student-T degrees of freedom (outlier robustness)} \\ \hline
    \end{tabular}
    \\
    \vspace{2mm}
    \parbox{14cm}{\small $^*$ Distributions are defined as $\mathcal{N}(\text{mean}, \text{standard deviation})$. Weakly informative priors centered at zero are adopted for the logistic regression coefficients ($\alpha, \beta_i$). For the slope parameters, centering at zero represents the null hypothesis that the covariates have no effect on the population split; that is, before seeing the data, we do not assume that $L_X$, $kT$, or $|Z|$ preferentially favor either channel. For the intercept, the prior $\alpha\sim\mathcal{N}(0,2)$ encodes a neutral prior on the baseline WD log-odds, so that in the absence of covariate effects the corresponding prior mixture probability is centered on $\pi=0.5$. The moderate prior widths provide soft regularization, so the coefficients move away from zero only when supported by the data, without imposing strong \textit{a priori} physical correlations.}
    \label{tab:priors}
    \end{center}
\end{table*}

To inform the prior probability of class membership, the mixture weights are modeled via logistic regression on the independent covariates:
\begin{equation}
    \pi_i(\Psi) = \mathrm{sigmoid}\left(\alpha + \beta_{L_X}\log_{10}L_{X,i} + \beta_{kT}kT_i + \beta_{|Z|}|Z_i|\right),
\end{equation}
where $\Psi = \{\alpha, \beta_{L_X}, \beta_{kT}, \beta_{|Z|}\}$ denotes the set of mixture-regression coefficients. Here, $L_{X,i}$ is the X-ray luminosity of source $i$ in units of erg s$^{-1}$, $kT_i$ is its thermal spectral temperature, and $|Z_i|$ is its absolute Galactic scale height. This parametrization allows spectral and environmental properties to inform the class-membership probabilities, while the primary spin-down information remains encoded in the likelihood through $x_i=\log_{10}(P_i\dot P_i).$ 

We denote $\Phi_c = \{\mu_c, \sigma_c, \sigma_{c,\mathrm{obs}}, \nu\}$ as the class-specific hyperparameters governing the Student-t densities $p_c(x_i | \Phi_c)$ defined in Eq. \eqref{student}. Marginalizing over the latent class assignments $c_i$, the per-source likelihood becomes:
\begin{equation}
    p(x_i | \Phi, \Psi) = \left(1 - \pi_i(\Psi)\right) p_{\mathrm{NS}}(x_i | \Phi_{\mathrm{NS}}) + \pi_i(\Psi) p_{\mathrm{WD}}(x_i | \Phi_{\mathrm{WD}}),
\end{equation}
with the full likelihood obtained by assuming conditional independence across sources.

\subsection{Inference and Model Evaluation}
Posterior inference is performed using the No-U-Turn Sampler (NUTS), a self-tuning Hamiltonian Monte Carlo algorithm \citep{duane/1987} implemented in \texttt{PyMC} \citep{abril-pla/2023}. We sampled 4 independent Markov chains, initializing each with 1,500 tuning steps followed by 3,000 recorded draws, yielding a total of 12,000 posterior samples. To ensure robust exploration of the complex hierarchical geometry, the target acceptance probability was set to 0.95.
Convergence is assessed using the Gelman-Rubin statistic ($\hat{R}$) \citep{vats/2021} and effective sample sizes.

\begin{figure}
   \centering
   \includegraphics[width=\hsize]{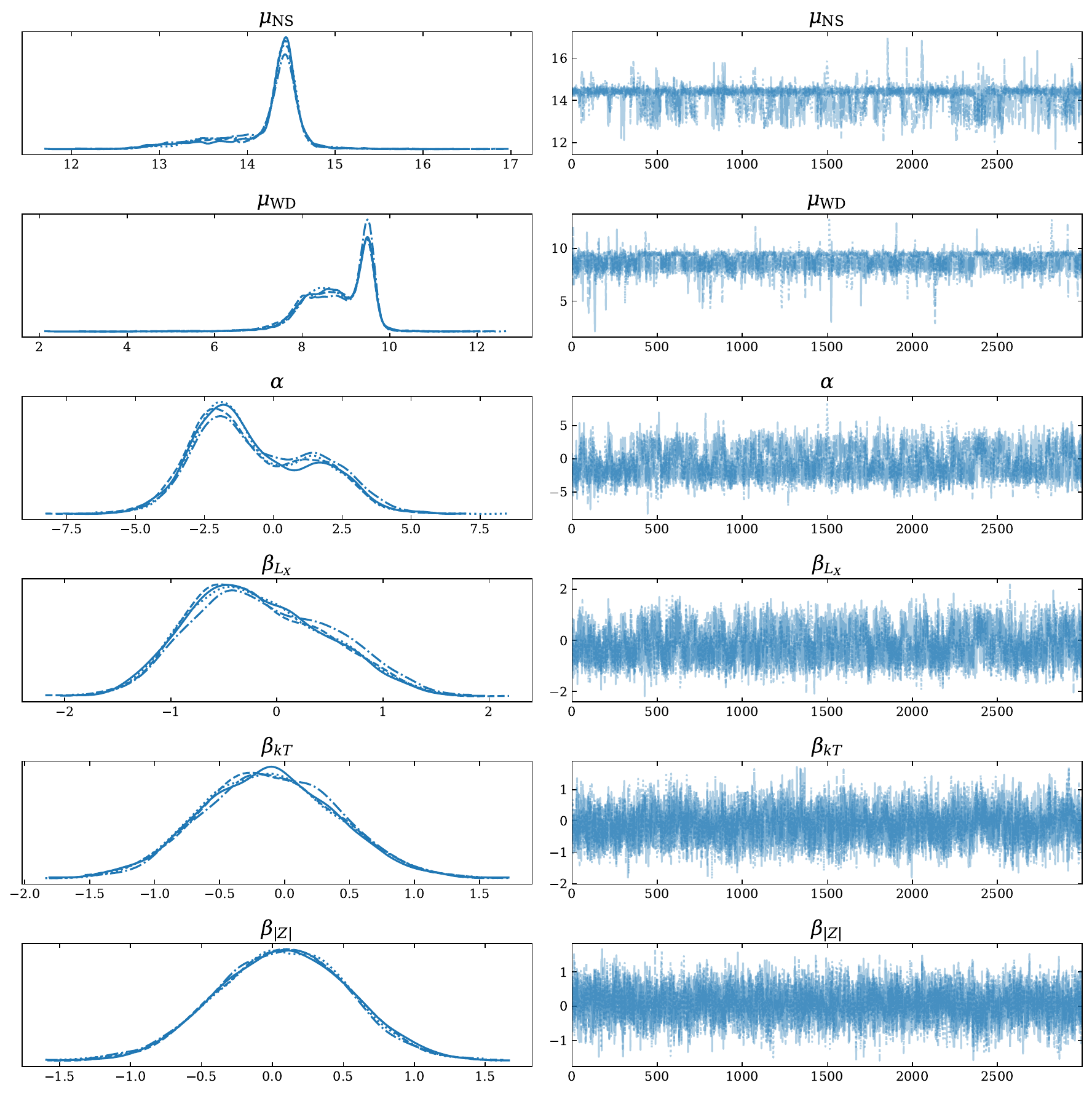}
   \caption{MCMC trace diagnostics for representative hyperparameters and mixture-regression coefficients. Left panels show marginal posterior densities; right panels show chain histories. The absence of long trends and the overlap of chains indicate good mixing and convergence. This supports the reliability of posterior summaries and model-comparison metrics.}
   \label{fig:trace}
\end{figure}

Trace plots for key hyperparameters and regression coefficients are shown in Fig.~\ref{fig:trace}. The right-column chain histories display the expected ``fuzzy caterpillar'' behavior, without long-range drifts, indicating stable exploration of posterior space. Trace plots indicate excellent mixing and convergence across 3,000 draws per chain (12,000 total).

Model comparison is carried out using Pareto-smoothed importance sampling leave-one-out cross-validation (PSIS-LOO) \citep{vehtari/2017}. Models are compared via the expected log predictive density (ELPD), reliability is assessed through the Pareto $k$ diagnostic, and we report $\Delta \mathrm{ELPD} \equiv \mathrm{ELPD}_{\mathrm{Mixture}} - \mathrm{ELPD}_{\mathrm{NS}}$, where NS denotes the NS-only null model.

\begin{figure}
   \centering
   \includegraphics[width=\hsize]{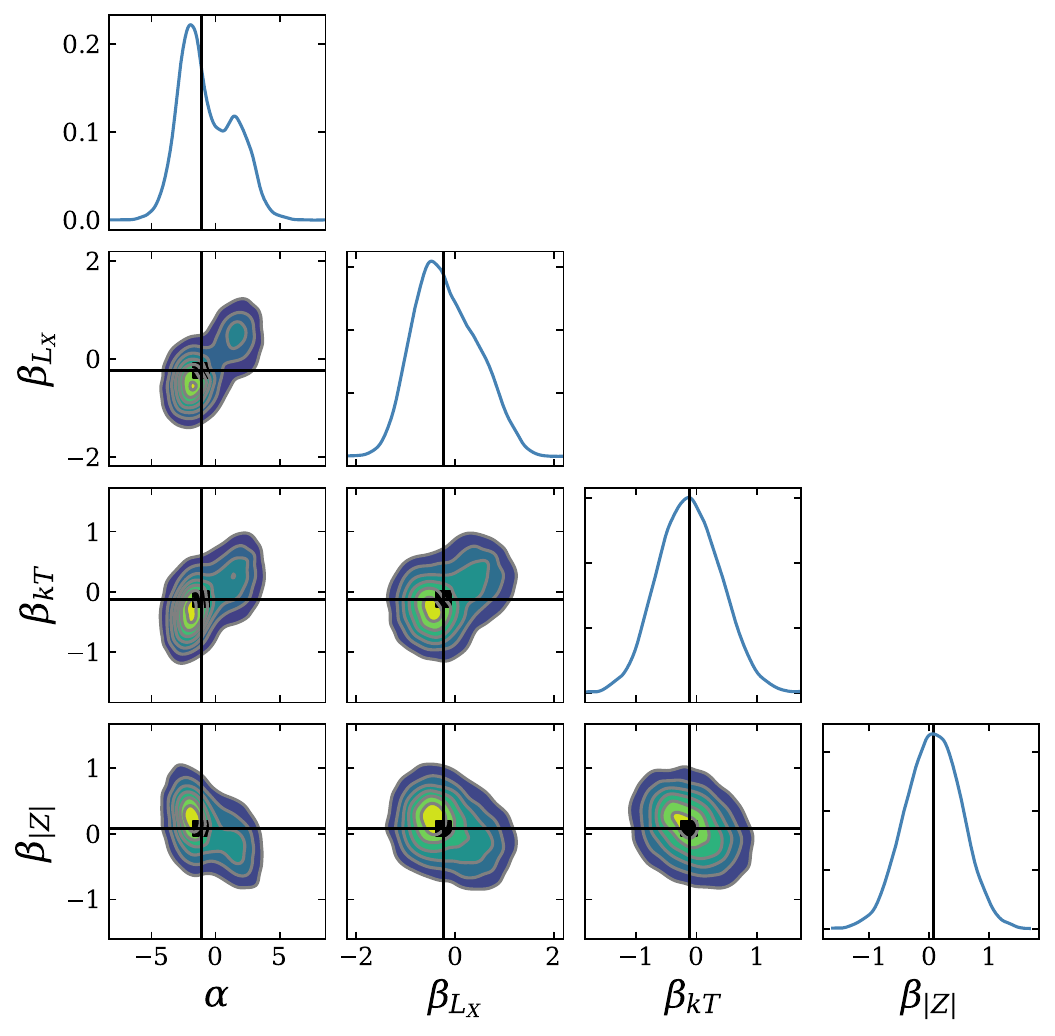}
   \caption{Corner plot of posterior samples for the logistic-mixture coefficients $(\alpha,\beta_{L_X},\beta_{kT},\beta_{|Z|})$. Diagonal panels show one-dimensional kernel density estimates (KDEs) of the marginal posterior distribution, with the vertical black lines indicating the posterior median values. Off-diagonal panels show the joint posterior structure and pairwise parameter correlations. The countour levels corresponde approximatelly to the $68\%$ and $95\%$ highest-density posterior regions. The posteriors remain broadly centered near zero but exhibit non-trivial correlations, indicating weak marginal evidence for individual covariate effects together with degeneracies in the mixing function.}
   \label{fig:corner}
\end{figure}

To evaluate covariate identifiability in the logistic mixing model, we inspect the joint posterior geometry of $(\alpha,\beta_{L_X},\beta_{kT},\beta_{|Z|})$ (Fig.~\ref{fig:corner}). The posterior distributions of the logistic-mixture coefficients are broadly consistent with their prior shapes and remain centered near zero, indicating that the current dataset does not provide strong evidence for a dependence of the mixture probabilities on the covariates. This suggests that, within the present sample, the assignment probabilities are largely insensitive to the included spectral and environmental variables. However, the joint posterior distributions exhibit non-trivial correlations between the coefficients, reflecting degeneracies in how different covariates can contribute to the mixing function. These correlations indicate that, although no single parameter strongly drives the classification, combinations of parameters can still influence the inferred probabilities.
This behavior is consistent with the limited size of the dataset, which restricts the ability to constrain higher-dimensional dependencies.

Posterior predictive checks are used to validate the adequacy of the model in reproducing the observed distribution of $x$; results are shown in Fig.~\ref{fig:ppc}.

\section{Results}

\subsection{Population-Level Parameters}

The population posterior distributions confirm that the magnetic-field hyperparameters of the neutron-star and white-dwarf components are clearly separated and tightly constrained. The the population joint posterior distributions are shown in \ref{app:full-corner}:
\begin{align}
\mu_{\mathrm{NS}} &\approx 14, \\
\mu_{\mathrm{WD}} &\approx 8-9,
\end{align}
These values are consistent with canonical expectations for neutron stars and highly magnetized white dwarfs, respectively. The additional scatter terms are also non-negligible, indicating that intrinsic dispersion beyond the mean component trends is required to account for the residual spread in the data. We report in Table \eqref{tab:posterior_summary} the posterior mean, standard deviation, and 95\% highest density intervals (HDI). The posterior medians are consistent with the reported means within uncertainties.

\begin{table}
\caption{Posterior summary statistics for all model parameters. We report the posterior mean, standard deviation SD, and 95\% highest density intervals (HDI)}
\centering
\begin{tabular}{lrrrrrrrrr}
 & Mean & SD & 2.5\% & 97.5\% \\
  \hline
$\mu_{\rm NS}$ & 14.202 & 0.486 & 12.969 & 14.764 \\
$\mu_{\rm WD}$ & 8.842 & 0.800 & 7.346 & 9.852 \\
$\sigma_{\rm NS}$ & 0.448 & 0.309 & 0.000 & 1.061 \\
$\sigma_{\rm WD}$ & 0.499 & 0.382 & 0.001 & 1.262 \\
$\sigma_{\rm NS, obs}$ & 0.259 & 0.186 & 0.000 & 0.610 \\
$\sigma_{\rm WD, obs}$ & 0.401 & 0.283 & 0.000 & 0.928 \\
$\alpha$ & -0.618 & 2.165 & -4.102 & 3.630 \\
$\beta_{L_X}$ & -0.143 & 0.638 & -1.292 & 1.086\\
$\beta_{kT}$ & -0.088 & 0.522 & -1.078 & 0.945 \\
$\beta_{|Z|}$ & 0.064 & 0.470 & -0.851 & 0.975 \\
$\nu$ & 10.630 & 9.869 & 0.001 & 30.308 \\
\end{tabular}
\label{tab:posterior_summary}
\end{table}

In contrast, the logistic-mixture coefficients remain broadly consistent with zero and show substantial posterior correlations. This indicates that the current dataset does not strongly constrain individual covariate effects. Instead, the inferred mixing probabilities are weakly informed by combinations of $L_X$, $kT$, and $|Z|$, reflecting degeneracies in the covariate space rather than a single dominant driver of class assignment.

\subsection{Model Comparison}

We compare the mixture model against a null hypothesis consisting of a single neutron star population. The difference in predictive performance is
\begin{equation}
\Delta \mathrm{ELPD} \equiv \mathrm{ELPD}_{\mathrm{Mixture}} - \mathrm{ELPD}_{\mathrm{NS}} \approx 1.3 \pm 1.6,
\end{equation}
indicating no statistically significant out-of-sample predictive preference for the mixture model over the NS-only model.

All Pareto $k$ values remain below 0.7, confirming the reliability of the LOO estimates.

\begin{figure}
   \centering
   \includegraphics[width=\hsize]{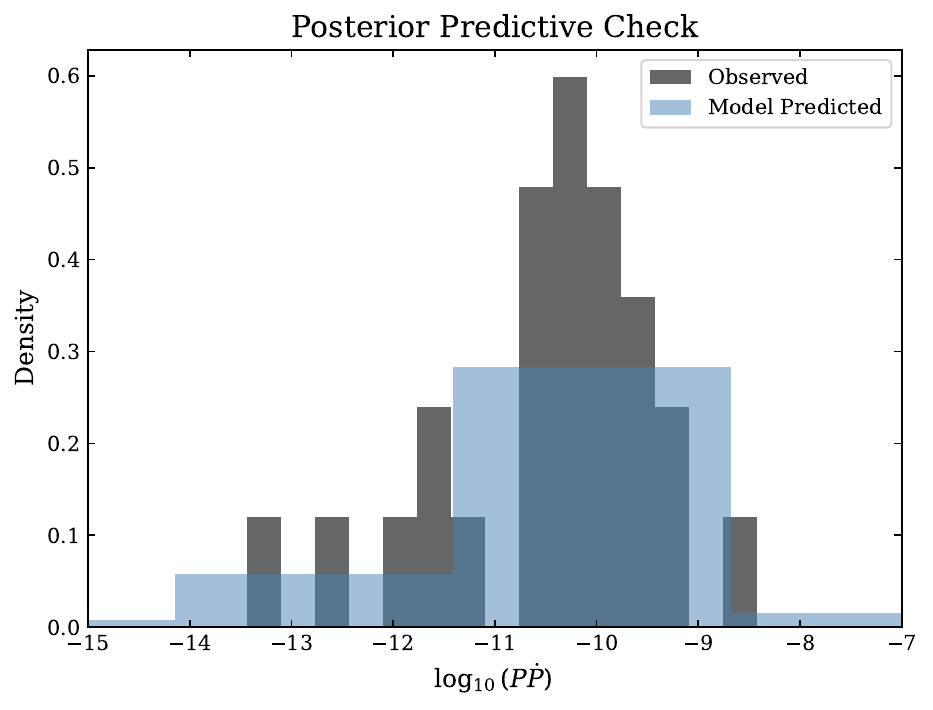}
   \caption{Posterior predictive check for the observable $x$. The broader predicted distribution reflects the WD component contribution at lower $x$, while the model correctly reproduces the dominant NS peak. The isolated observed bins near $x = -13$ to $-14$ correspond to the low spin-down outliers.}
   \label{fig:ppc}
\end{figure}

\begin{figure*}[t]
   \centering
   \includegraphics[width=0.48\textwidth]{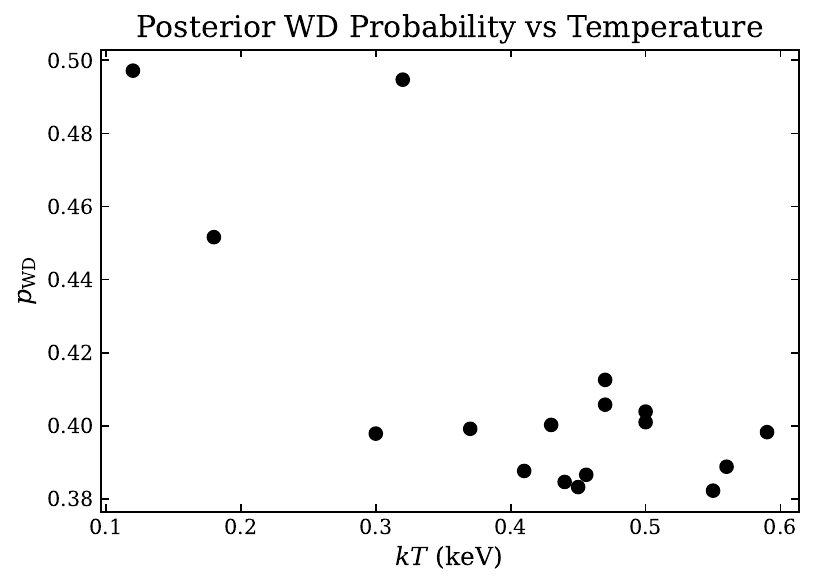}\hfill
   \includegraphics[width=0.48\textwidth]{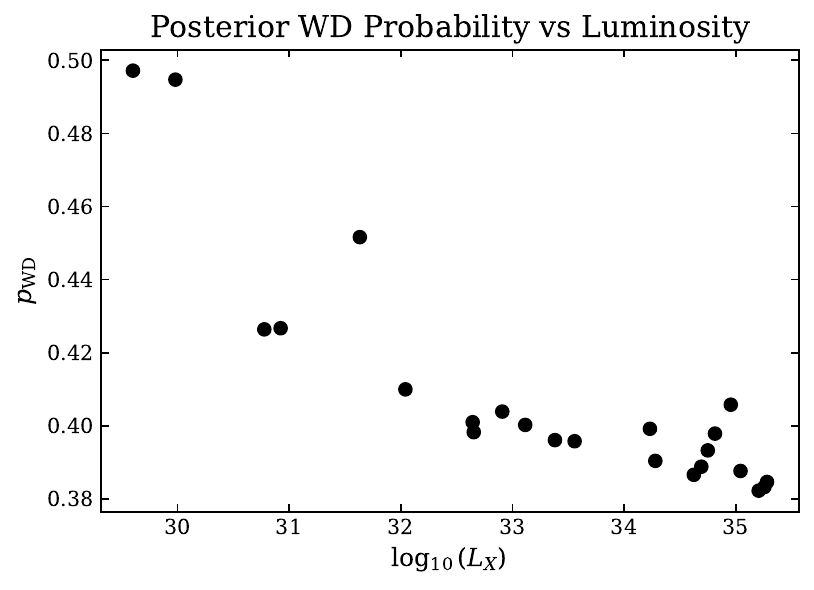}
   \caption{Posterior white-dwarf membership probability as a function of surface temperature $kT$ (left) and X-ray luminosity $L_X$ (right). In both panels, the inferred probabilities remain broadly flat across the observed covariate ranges, supporting the conclusion that these covariates provide limited discriminatory power in the current sample.}
   \label{fig:pwd-covariates}
\end{figure*}

Posterior predictive validation is shown in Fig.~\ref{fig:ppc}, where the model-generated distribution closely tracks the observed histogram of $x$. This agreement indicates that the hierarchical mixture captures the main features of the observed spin-parameter distribution. It is important to clarify the structural role of the multi-wavelength parameters in this hierarchy. While the posterior predictive check, Fig.~\ref{fig:ppc}, confirms the model's ability to reproduce the primary timing observable $x$, the parameters $L_X$, $kT$, and $|Z|$ are strictly covariates. Because they act as fixed independent inputs to the logistic mixing prior rather than generative outputs of a class-conditional likelihood, a posterior predictive check for these specific variables is mathematically inapplicable. Their physical consistency within the model is instead evaluated by their resulting posterior dependencies. We further examine the posterior WD probabilities as a function of the covariates $L_X$ and $kT$, Fig.~\ref{fig:pwd-covariates}. Although no statistically significant dependence is observed, the left panel of Fig. \ref{fig:pwd-covariates} suggests a mild tendency for cooler sources (lower $kT$) to exhibit somewhat larger posterior WD probabilities. Given the limited sample size and the broad posterior uncertainties of the logistic-regression coefficients, this apparent trend should be interpreted cautiously, as it may be driven by a small number of objects rather than reflecting a robust population-level correlation. This confirms that the covariates do not provide strong discriminatory power within the current dataset, consistent with the posterior distributions of the logistic-mixture coefficients.

\subsection{Posterior Classification}

Posterior probabilities for individual objects show that several low-$\dot{P}$ magnetars occupy an intermediate regime, with
\begin{equation}
p_{\mathrm{WD},i} \sim 0.4\text{-}0.5.
\end{equation}

Notably, sources such as SGR~0418+5729 and Swift~J1822.3$-$1606 exhibit the highest probabilities, although none are decisively classified as white dwarfs.

\begin{figure}
   \centering
   \includegraphics[width=\hsize]{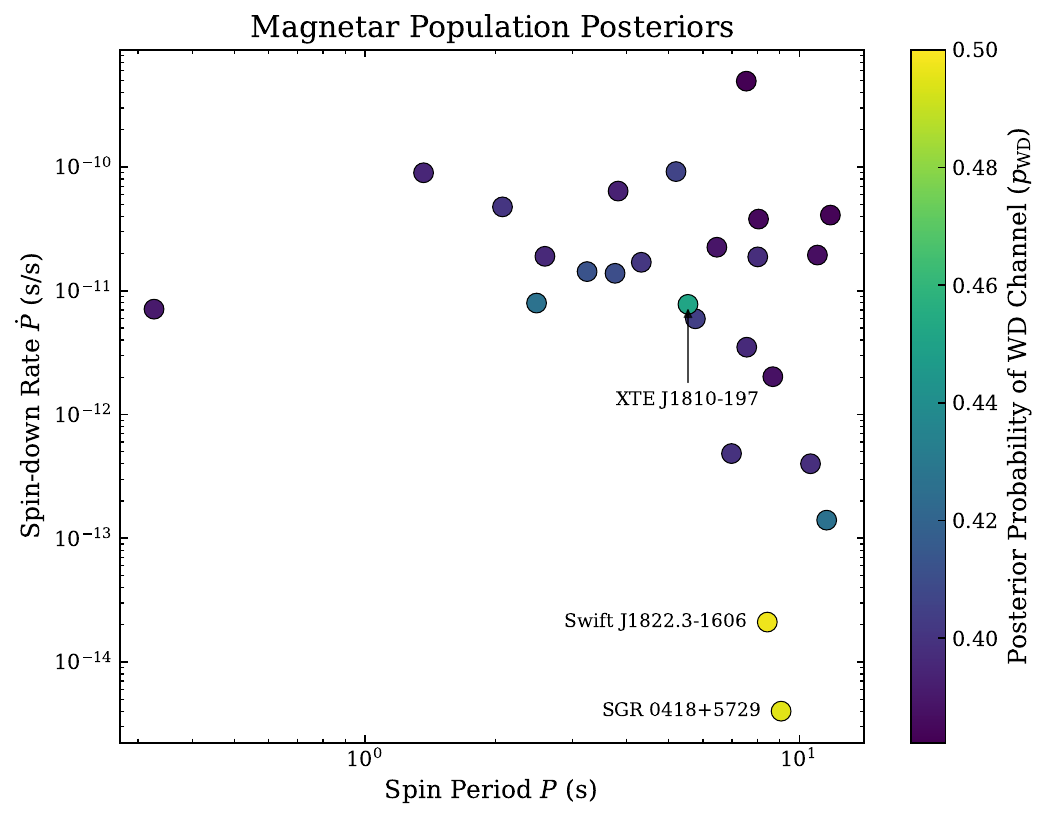}
   \caption{$P$-$\dot{P}$ diagram color-coded by posterior white-dwarf membership probability, $p_{\mathrm{WD},i}$. The map highlights candidate transitional sources with comparatively high posterior probability under the WD channel. These sources appear as outliers of the neutron-star locus rather than as a detached second branch.}
   \label{fig:ppdot-posterior}
\end{figure}

Figure~\ref{fig:ppdot-posterior} presents the $P$-$\dot{P}$ plane color-coded by posterior WD probability, $p_{\mathrm{WD},i}$. Most sources remain tightly concentrated in the neutron-star-dominated region, while low spin-down objects (including SGR~0418+5729 and Swift~J1822.3$-$1606) separate from the main cluster and exhibit elevated $p_{\mathrm{WD},i}$, consistent with a transitional/outlier interpretation rather than a clearly distinct second population.

\section{Discussion}

\subsection{Implications for Magnetar Populations}

The absence of strong statistical evidence for a second population suggests that the current dataset is consistent with a single neutron star population. This interpretation is consistent with the small model-comparison gain ($\Delta \mathrm{ELPD} \approx 1.3 \pm 1.6$) and with Fig.~\ref{fig:ppc}, where the posterior predictive distribution reproduces the observed $x$ histogram without requiring a strongly separated second component.

At the same time, the presence of objects with intermediate posterior probabilities indicates deviations from standard magnetar behavior. In that sense, the exploratory split highlighted in Figs.~\ref{fig:eda-pca} and \ref{fig:eda-rf} is better interpreted as evidence for a transitional/outlier tail than as definitive evidence for a separate compact-object class.

\subsection{Interpretation of Ambiguous Objects}

Low-$\dot{P}$ sources such as SGR~0418+5729 have previously been identified as outliers in the magnetar population. Our Bayesian results indicate that these objects lie in a transitional regime in parameter space, yielding the highest posterior probabilities for the white-dwarf channel within the sample. This behavior is visualized in Fig.~\ref{fig:ppdot-posterior}, where they occupy the edge of the neutron-star-dominated locus rather than forming a clearly detached white-dwarf branch.

To test the physical plausibility of that classification, we estimate the characteristic thermal emitting radius from blackbody scaling. Here, $L_X$ is the X-ray luminosity, $R_{\mathrm{emit}}$ the characteristic emitting radius, $\sigma_{\mathrm{SB}}$ the Stefan--Boltzmann constant, $T$ the effective blackbody temperature, and $k_B$ the Boltzmann constant.
\begin{equation}
L_X = 4\pi R_{\mathrm{emit}}^2\sigma_{\mathrm{SB}}T^4,
\end{equation}
and therefore
\begin{equation}
R_{\mathrm{emit}} = \sqrt{\frac{L_X}{4\pi\sigma_{\mathrm{SB}}T^4}},
\end{equation}
with $T = (kT)/k_B \simeq 1.160\times10^7\,(kT/\mathrm{keV})\,\mathrm{K}$. Throughout this estimate, we assume isotropic blackbody emission and treat the quoted $L_X$ as a proxy for bolometric luminosity; if $L_X$ is band-limited, $R_{\mathrm{emit}}$ should be interpreted as a lower-limit scale.

Using representative values for SGR~0418+5729 ($L_X \approx 9.6\times10^{29}\,\mathrm{erg\,s^{-1}}$, $kT=0.32\,\mathrm{keV}$) \citep{rea/2010, rea/2012, olausen/2014}, we obtain $T\approx3.7\times10^6\,\mathrm{K}$ and
\begin{equation}
R_{\mathrm{emit}} \approx 2.7\times10^3\,\mathrm{cm} \approx 2.6\times10^{-2}\,\mathrm{km} \approx 26\,\mathrm{m}.
\end{equation}

For a massive white dwarf with $R_{\mathrm{WD}}\approx3000\,\mathrm{km}$, the implied active area fraction is
\begin{equation}
f_A = \left(\frac{R_{\mathrm{emit}}}{R_{\mathrm{WD}}}\right)^2 \approx 7.9\times10^{-11}\;\; (\approx 7.9\times10^{-9}\%).
\end{equation}
Such a small emitting region is not impossible (e.g., localized polar-cap heating), but it imposes a stringent geometric constraint on a white-dwarf interpretation. By contrast, this scale is more naturally accommodated on a $\sim10\,\mathrm{km}$ neutron-star surface. Consequently, while timing-based statistics can shift these outliers toward the WD-like channel, thermal-emission geometry appears more consistent with a neutron-star origin.

The X-ray luminosities reported in the McGill Magnetar Catalog are typically derived from spectral fits that include thermal components, often interpreted as emission from hot regions on the stellar surface. In the neutron-star framework, such emission is naturally explained by magnetic-field decay and crustal heating.

In contrast, the interpretation of these luminosities within a white-dwarf scenario is less straightforward. The larger radii and lower magnetic-energy densities of white dwarfs make it more difficult to sustain localized high-temperature regions through magnetic dissipation alone. Alternative explanations, such as accretion from a companion, could in principle contribute to the observed luminosities, but they would require persistent or transient binary interactions that are not generally supported by observations of magnetar sources. Moreover, reconciling magnetar-like bursting activity with an accreting white-dwarf scenario remains challenging. These considerations suggest that, while the statistical model allows for intermediate classifications, the physical interpretation of such objects as white dwarfs faces significant constraints from their X-ray properties.

Notably, the clustering algorithm and Bayesian posterior also highlight XTE J1810-197. As the first known transient magnetar, its low quiescent X-ray luminosity and cooler surface temperature naturally shift it toward the anomalous parameter space, demonstrating that our data-driven pipeline successfully recovers physically distinct sub-classes within the McGill catalog.

\subsection{Limitations and Future Work}

The primary limitation of this study is the small sample size, which restricts the ability to distinguish between competing population models. Although inference diagnostics indicate stable sampling and identifiable posterior structure (Figs.~\ref{fig:trace} and \ref{fig:corner}), future observations (particularly improved measurements of spin-down rates and distances) will be critical in refining these constraints. In addition, timing noise and possible upper limits in some $\dot{P}$ measurements can propagate nonlinearly into $x=\log_{10}(P\dot{P})$, and should be modeled explicitly in expanded future samples.

Extending this framework to larger samples or incorporating additional observables may provide stronger discriminatory power between competing formation channels.

\section{Conclusion}

We presented a joint data-driven and physics-informed analysis to test whether the observed magnetar sample is better described by a single neutron-star population or by a mixture that includes a white-dwarf-like channel. The workflow combines unsupervised structure discovery in the five-dimensional feature space $(P,\dot{P},L_X,kT,|Z|)$, supervised feature-ranking diagnostics, and a hierarchical Bayesian mixture model with covariate-dependent class probabilities.

Although exploratory machine-learning diagnostics reveal a reproducible sub-structure (including known low-$\dot{P}$ outliers), Bayesian model comparison does not yield statistically significant support for a distinct second population: $\Delta\mathrm{ELPD}\approx 1.3\pm1.6$ remains compatible with the NS-only hypothesis. At the same time, posterior source-level probabilities indicate that a few objects occupy an intermediate regime rather than a sharply detached branch.

Our physical consistency check using blackbody scaling reinforces this interpretation. For representative SGR~0418+5729 parameters, the inferred emitting radius is of order tens of meters, implying an extremely small active-area fraction if interpreted on a $\sim3000$ km white dwarf, while remaining naturally compatible with localized emission on a neutron-star surface. Taken together, the statistical and phenomenological evidence favors a predominantly neutron-star magnetar population with a transitional/outlier tail.

The main limitation is sample size and measurement quality, especially for $\dot{P}$ in low spin-down sources. Future progress will come from enlarged magnetar catalogs, improved timing and distance constraints, and richer multi-wavelength characterization (including better bolometric luminosity estimates). Extending the hierarchical model to treat censoring/upper limits and additional observables should provide stronger discrimination between competing formation channels.

\section*{Acknowledgments}
 RVL was supported by INCT-FNA (Instituto Nacional de Ci\^encia e Tecnologia, F\'{\i}sica Nuclear e Aplica\c c\~oes), research Project No.~464898/2014-5, and acknowledges support from CAPES/CNPq.

\section*{Data Availability}
The codes used in this study are available from the corresponding author upon reasonable request. All datasets used in this study are publicly available and can be accessed from their original sources:
\begin{itemize}
    \item McGill Online Magnetar Catalog: \url{https://www.physics.mcgill.ca/~pulsar/magnetar/main.html}
\end{itemize}

\appendix
\clearpage
\onecolumn

\section{Population Corner Plot}
\label{app:full-corner}

\begin{figure}[!ht]
   \centering
   \includegraphics[width=\textwidth]{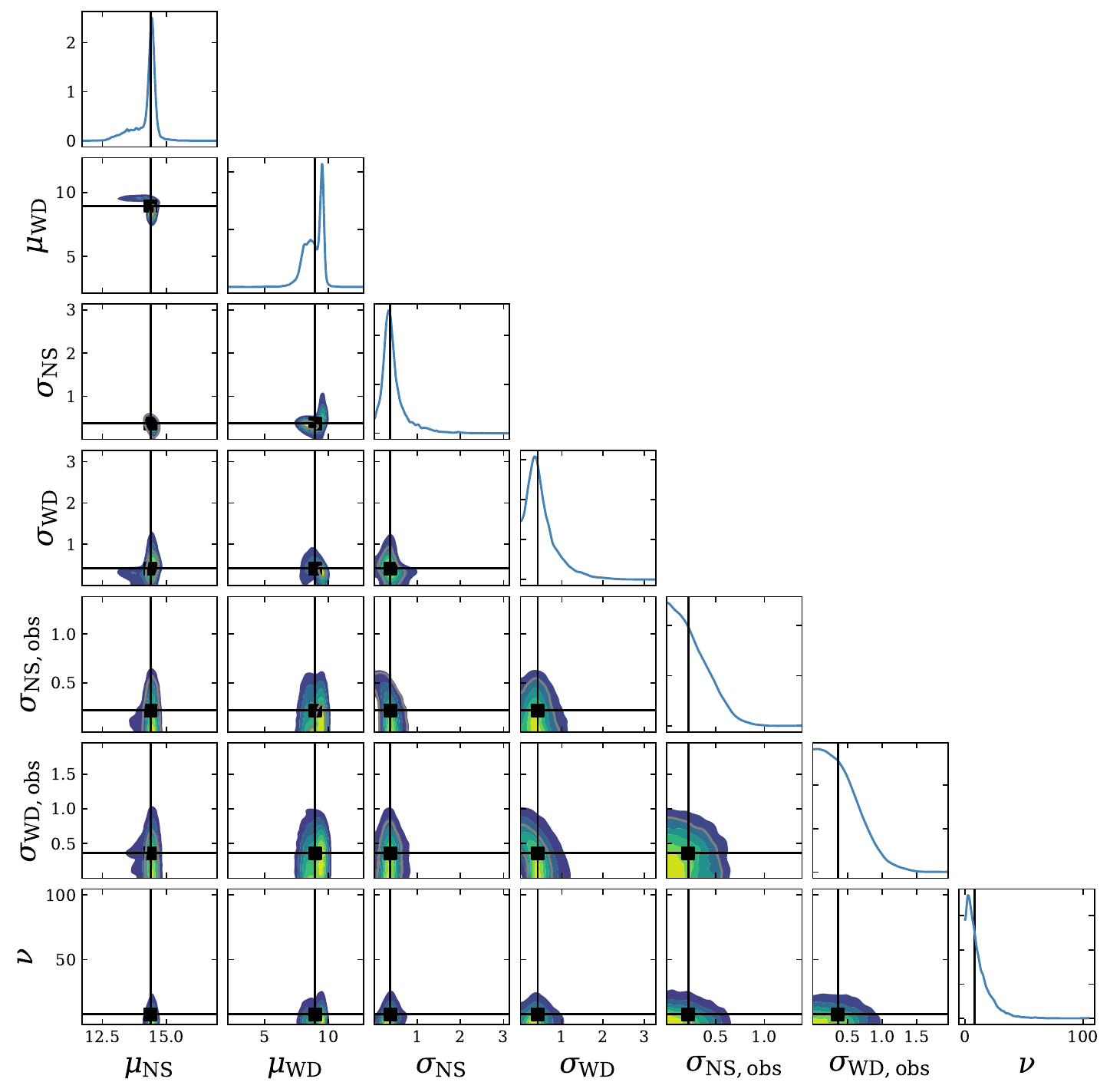}
   \caption{Corner plot of the posterior samples for the population hyperparameters of the hierarchical Bayesian mixture model. Diagonal panels show the one-dimensional marginal posterior densities for each sampled parameter, with the vertical black lines indicating the posterior median values. Off-diagonal panels show pairwise posterior correlations and kernel-density credible regions derived from the posterior samples. The contour levels correspond approximately to the $68\%$ and $95\%$ highest-density posterior regions.}
   \label{fig:full-corner}
\end{figure}

\clearpage
\twocolumn

\bibliographystyle{elsarticle-harv}
\bibliography{references}






\end{document}